\pdfoutput=1
\documentclass[rnote]{aa} 

\usepackage{graphicx}
\usepackage{txfonts}
\def\14{\rm 1.4\,GHz}
\def\27{\rm 2.7\,GHz}
\def\whz1{$\,\rm W\,Hz^{-1}$}
\def\kms1{$\,\rm km\,s^{-1}$}

\def\gsim{\mathrel{\lower0.6ex\hbox{$\buildrel {\textstyle >}
\over {\scriptstyle \sim}$}}}

\def\lsim{\mathrel{\lower0.6ex\hbox{$\buildrel {\textstyle <}
\over {\scriptstyle \sim}$}}}

\begin{document}


\title{Cosmic downsizing of powerful radio galaxies to low radio luminosities} 

   \author{E. E. Rigby\inst{1},
          J. Argyle\inst{1,2},
          P. N. Best\inst{3},
          D. Rosario\inst{4}
          \and
          H. J. A. R\"{o}ttgering\inst{1}
          }

   \institute{Leiden Observatory, Leiden University, P.O. Box 9513, 2300 RA Leiden, The Netherlands\\
              \email{emmaerigby@gmail.com}
         \and
             School of Physics and Astronomy, University of St Andrews, North Haugh, St Andrews, KY16 9SS, UK
         \and
             SUPA, Institute for Astronomy, Royal Observatory Edinburgh, Blackford Hill, Edinburgh EH9 3HJ, UK
         \and
             Max Planck Institute for Extraterrestrial Physics, Giessenbachstrasse 1, 85748 Garching, Germany
             }

\abstract
{}
{At bright radio powers ($P_{\rm 1.4 GHz} > 10^{25}$ W/Hz) the space density of the most powerful sources peaks at higher redshift than that of their weaker counterparts. This paper establishes whether this luminosity--dependent evolution persists for sources an order of magnitude fainter than those previously studied, by measuring the steep--spectrum radio luminosity function (RLF) across the range $10^{24} < P_{\rm 1.4 GHz} < 10^{28}$ W/Hz, out to high redshift. }  
{A grid--based modelling method is used, in which no assumptions are made about the RLF shape and high--redshift behaviour. The inputs to the model are the same as in \citet{paper1}: redshift distributions from radio source samples, together with source counts and determinations of the local luminosity function. However, to improve coverage of the radio power vs. redshift plane at the lowest radio powers, a new faint radio sample is introduced. This covers 0.8 sq. deg., in the Subaru/XMM--Newton Deep Field, to a 1.4 GHz flux density limit of $S_{\rm 1.4 GHz} \geq 100~\mu$Jy, with 99\% redshift completeness. }
{The modelling results show that the previously seen high--redshift declines in space density persist to $P_{\rm 1.4 GHz} < 10^{25}$ W/Hz. At $P_{\rm 1.4 GHz} > 10^{26}$ W/Hz the redshift of the peak space density increases with luminosity, whilst at lower radio luminosities the position of the peak remains constant within the uncertainties. This `cosmic downsizing' behaviour is found to be similar to that seen at optical wavelengths for quasars, and is interpreted as representing the transition from radiatively efficient to inefficient accretion modes in the steep--spectrum population. This conclusion is supported by constructing simple models for the space density evolution of these two different radio galaxy classes; these are able to successfully reproduce the observed variation in peak redshift.}
{}


\authorrunning{Rigby et al.}

\maketitle

\section{Introduction}

Radio--loud active galactic nuclei are a key component driving galaxy evolution; the feedback their expanding radio jets provide is essential for preventing large--scale cluster cooling flows and halting the growth of massive elliptical galaxies \citep[e.g.][]{fabian06, best06, best07, croton06, bower06}. To understand the timescales upon which these processes occur, it is important to first understand the evolution of the radio luminosity function (RLF) to high--redshift. An early measurement of this came from \citet{DP90}, who found increases in the space density of both flat and steep--spectrum radio AGN of two to three orders of magnitude, over that seen locally. They also saw the first indication of an expected higher redshift density decline at $z \sim 2.5$, corresponding to the build--up of these objects in the early Universe. However, their work, and that of subsequent studies \citep[e.g.][]{shaver, jarvis01b, waddington01}, lacked the necessary depth and volume needed to unambiguously measure this high--redshift behaviour. 

This situation improved with the development of the Combined EIS--NVSS Survey of Radio Sources \citep[CENSORS;][]{CENSORS1}: a survey designed to maximise the coverage of steep--spectrum radio sources close to the high--redshift break in the RLF. \citet[][hereafter R11]{paper1} used CENSORS, combined with additional radio source samples, source counts and determinations of the local RLF, to investigate the space density evolution of the $P_{\rm 1.4 GHz}>10^{25}$ W/Hz steep--spectrum population via grid--based modelling with no prior assumptions included about the high--redshift behaviour. This robustly identified the post--peak space density decline in the RLF, and found that this turnover appears to be luminosity--dependent; at lower radio powers ($P_{\rm 1.4GHz} = 10^{25-26}$ W/Hz) the space densities peak at $z \gsim 1$, but the peak moves to higher redshift for the more luminous objects ($z \gsim 3$ for $P_{\rm 1.4 GHz} > 10^{27}$ W/Hz). A luminosity dependence in the position of the steep--spectrum RLF peak can be interpreted as a sign of `cosmic downsizing', in which the most massive black holes form at earlier epochs than their less massive counterparts. This has also been seen for other AGN populations, selected at other radio, optical, far--infrared and X--ray wavelengths \citep[e.g.][]{zotti09, hasinger05, richards05, mcalpine13, delvecchio14}, as well as reproduced in simulations of black hole growth \citep[e.g.][]{fanidakis12, hirschmann12, hirschmann13}. 

Steep--spectrum radio sources can be split into two distinct populations: typically powerful objects with `standard' accretion of cold gas and likely to be merger driven (`cold--mode'); and typically weaker sources with radiatively inefficient accretion directly from their hot gas haloes (`jet--mode' or `hot--mode'). Locally, the latter dominate below $P_{\rm 1.4 GHz} \sim 10^{26}$ W/Hz, but both classes are present at all powers \citep{best2012}. The major limitation of R11 was the lack of constraint at lower radio powers ($P \lesssim 10^{25}$ W/Hz) out to $z \gtrsim 2$. The R11 power range was thus only able to probe the RLF regimes where the cold--mode sources are in the majority, and therefore could not draw any conclusions about the relative contributions of the two classes across cosmic time. This paper builds and expands on this initial foundation by including radio powers an order of magnitude fainter than R11, allowing the RLF evolution to be measured across $24 \leq \log P_{\rm 1.4 GHz} \leq 28$, to high redshift. This improved coverage means that the apparent luminosity--dependence of the peak space density can be investigated for the first time from the brightest sources, down to the regime where the dominant steep--spectrum population is shifting between the two types.

The layout of the paper is as follows: Section \ref{mod_proc} describes the modelling method used to determine the best--fitting steep spectrum RLFs; Section \ref{sxdf_sec} presents the SXDF radio source sample; the results of incorporating this into the modelling process are given in Section \ref{mod_res}; Section \ref{z_turn} discusses the luminosity--dependence of the position of the RLF peak; finally the results are summarised and interpreted in Section \ref{summ}. 
Throughout this paper values for the cosmological parameters of H$_{0} = 70$~km~s$^{-1}$Mpc$^{-1}$, $\Omega_{\rm m} = 0.3$ and $\Omega_{\rm \Lambda} = 0.7$ are used and the radio spectral index, $\alpha$ is defined as $S_{\nu} \propto \nu^{-\alpha}$.

\section{Modelling process}
\label{mod_proc}

\begin{figure}
\centering
\includegraphics[scale=0.36, trim=80 100 40 100, clip]{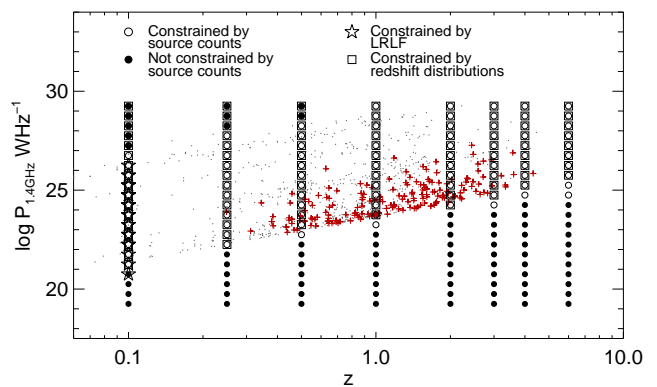}
\caption{\protect\label{pz} The distribution of grid points in the $P$--$z$ plane with the constraints offered by the three different types of input dataset highlighted. Grey dots show the positions of the individual sources in the four original radio source samples taken from R11. The sources in the new SXDF sample are highlighted with red crosses.}
\end{figure}

The modelling method in R11 makes no assumptions about the shape and evolution of the RLF. It takes as input five radio galaxy redshift distributions, measurements of the radio source counts over 0.05 mJy $\leq S_{1.4 GHz} \leq$ 94 Jy, and determinations of the local radio luminosity function. The radio samples are selected such that they provide a good coverage of the radio luminosity -- redshift plane out to $z \sim 4$ (Figure \ref{pz}). 

These input data are used to constrain co--moving space densities, $\rho$, determined at various points on a $P$--$z$ grid of radio luminosities and redshifts, from which the cosmic evolution of the radio galaxy population can simply be read directly. This grid is made up of three components -- steep--spectrum, flat--spectrum and star--forming sources -- representing the different radio source populations present. Since the steep--spectrum component is the dominant population at the luminosities and redshifts considered in R11, it is the only one allowed to vary in the modelling process (see Section \ref{sxdf_sec} for further discussion). The constant star--forming grid is created by evolving the local star--forming galaxy luminosity function of \citet{Sadler02} by $P/(1+z)^{2.5}$ to $z = 2$ and by $0.06P$ at $z>2$; the constant flat--spectrum grid is taken as the median of the set of evolutionary models presented in \citet{DP90} for this source population. The starting point for the steep--spectrum grid is formed by evolving the local AGN RLF \citep{Sadler02} by $(1+z)^{3}$ in density. 

The grids cover a range of $19.25 \leq \log P_{\rm 1.4 GHz} \leq 29.25$ in radio luminosity, equally spread in steps of $\Delta \log P_{\rm 1.4 GHz} = 0.5$, and $0.1 \leq z \leq 6.0$ in redshift, evaluated at $z=0.1,0.25,0.5,1.0,2.0,3.0,4.0,6.0$. The space density at any intermediate $P$--$z$ grid point can simply be interpolated from its 4 neighbours. Figure \ref{pz} illustrates the distribution of grid points, together with the corresponding constraints offered by the input data. Unconstrained densities are excluded from the fitting process. 

The best--fitting space density grid is determined using the \verb1amoeba1 algorithm for downhill simplex minimisation \citep{downhillsimplex}, run in a multi--stage loop with varying scaling and tolerance parameters for the initial steps. The goodness of fit for a particular model grid is the combined likelihood of the predicted source counts, redshift samples and local luminosity functions, compared to the real data. The uncertainties in the model grid are taken as the marginalized error, calculated using the inverse Hessian matrix. 

Full details of both the modelling process and the input datasets used can be found in R11. 

\section{New input data: the Subaru/XMM--Newton Deep Field radio sample}
\label{sxdf_sec}

The faintest radio galaxy sample used in R11 was the AGN subsample of the VLA--COSMOS survey \citep{smolcic}, with a 1.4 GHz flux density limit of 100 $\mu$Jy. This dataset was only complete to $z \leq 1.3$ which, since the next faintest sample had a 2 mJy flux density limit, severely limited the high--redshift constraint in the $P$--$z$ grid at low radio powers. This, therefore, restricted the RLF evolution analysis to $\log P_{\rm 1.4 GHz} \geq 25$ only. To improve this constraint a radio survey of comparable depth but better redshift coverage, carried out in the Subaru/XMM--Newton Deep Field \citep[SXDF;][]{simpson06, simpson12}, is now included in the modelling. This covers 0.81 sq. deg. and contains a complete sample of 505 sources with $S_{\rm 1.4 GHz} \geq 100~\mu$Jy. Robust redshifts, both spectroscopic (51\%) and photometric, exist for 99\% of the sample. The photometric redshifts for the remaining 7 sources are contaminated by foreground objects and are therefore excluded here.

At this flux density level $\gtrsim$50\% of the sources are expected to be star--forming galaxies \citep[e.g.][]{norris}. Since this paper is concerned with the behaviour of the steep--spectrum population (which comprises the majority of the remainder), the catalogue is restricted to radio--loud AGN only, as described below. In principle, this step is not necessary as the evolution of the star--forming RLF in the modelling process should be accounted for by the treatment applied to the star--forming grid. However, uncertainties in the evolution parameters used for this, combined with the non--negligible contribution of this population in this survey, means that it is prudent to perform this step to limit the effect of any errors.

The AGN/star--formation separation is done by calculating the ratio of mid--infrared, $S_{\rm 24 \mu m}$, to radio flux density via the k--corrected $q$ parameter: $q = \log(S_{\rm 24 \mu m} / S_{\rm 1.4 GHz})$; following \citet{donley05}, AGN are defined as sources with $q<0$, indicating excess radio emission over that from star--formation. Applying this cut results in a final sample of 179 sources. The $q$ parameter adopted is higher than the $q = -0.23$ value used previously to select radio--loud objects \citep[e.g.][]{ibar08}. However, inspection of Figure 6 in \citet{simpson12} shows that the higher cut is more appropriate for this sample. In practice, the difference in the modelling results from changing $q$ is negligible within the errors. 

The AGN cut also removes radio quiet quasars from the sample. These also begin to be a larger fraction of the radio population at $S_{\rm 1.4 GHz} < 1$ mJy, and display radio emission which may arise mainly from star--formation instead of black hole accretion, though this is still a topic of active debate \citep[][and references therein]{smolcic15}. Their contribution to the overall population is not explicitly included here, but rather assumed to be covered by the star--forming grid. 

\begin{figure}
\centering
\includegraphics[scale=0.37, trim=80 70 40 40, clip]{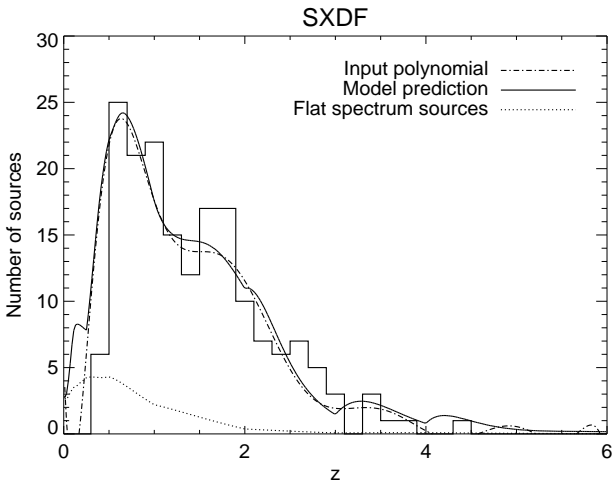}
\caption{\protect\label{z_sxdf} The redshift distribution histogram for the SXDF catalogue after removing star--forming galaxies. The solid line shows the best--fit model prediction for this sample, and the dot--dash line shows the 12th order polynomial used to represent the distribution in the modelling process. The contribution to the best--fit by the flat--spectrum grid is also shown (dotted line) to illustrate the dominance of the steep--spectrum grid at all redshifts.}
\end{figure}

The redshift distribution of the sample is shown in Figure \ref{z_sxdf}, together with the 12th order polynomial used to represent it in the modelling process. This step is necessary as the low number of sources at high--redshift makes a direct calculation of the $\chi^{2}$ difficult (see R11 for further details). 

The VLA--COSMOS sample previously used in the modelling process is now replaced here by the SXDF sample described above. 

\section{Model results}
\label{mod_res}

\subsection{Dataset comparison}

The independent redshift distribution generated from the best--fit steep--spectrum $P$--$z$ grid is shown in Figure \ref{z_sxdf}; this agrees well with the real dataset at $z>0.3$. Below this the SXDF potentially misses resolved sources \citep{simpson12}, which may explain the apparent overprediction of the model at low redshift. The Figure also shows the small relative contribution to the total population from the un--fitted flat--spectrum grid.

The total number of AGN (i.e. flat-- plus steep--spectrum sources) given by the model is $199 \pm 31$ ($188 \pm 29$ at $z \geq 0.3$), which compares well to the actual figure of 179 (178 at $z \geq 0.3$). In addition, the predicted number of star--forming galaxies (377) broadly agrees with the 326 observed in the SXDF sample and suggests that the treatment of this population in the modelling is sufficient to not give rise to additional uncertainty in the grid results. 

The agreement between the model and the other input datasets is also good, however these comparisons are not explicitly shown here as they are essentially unchanged from the results previously presented in R11.

\subsection{Model RLFs}
\label{rlf_sec}

\begin{figure*}
\centering
\includegraphics[scale=0.3, trim=40 80 40 80, clip]{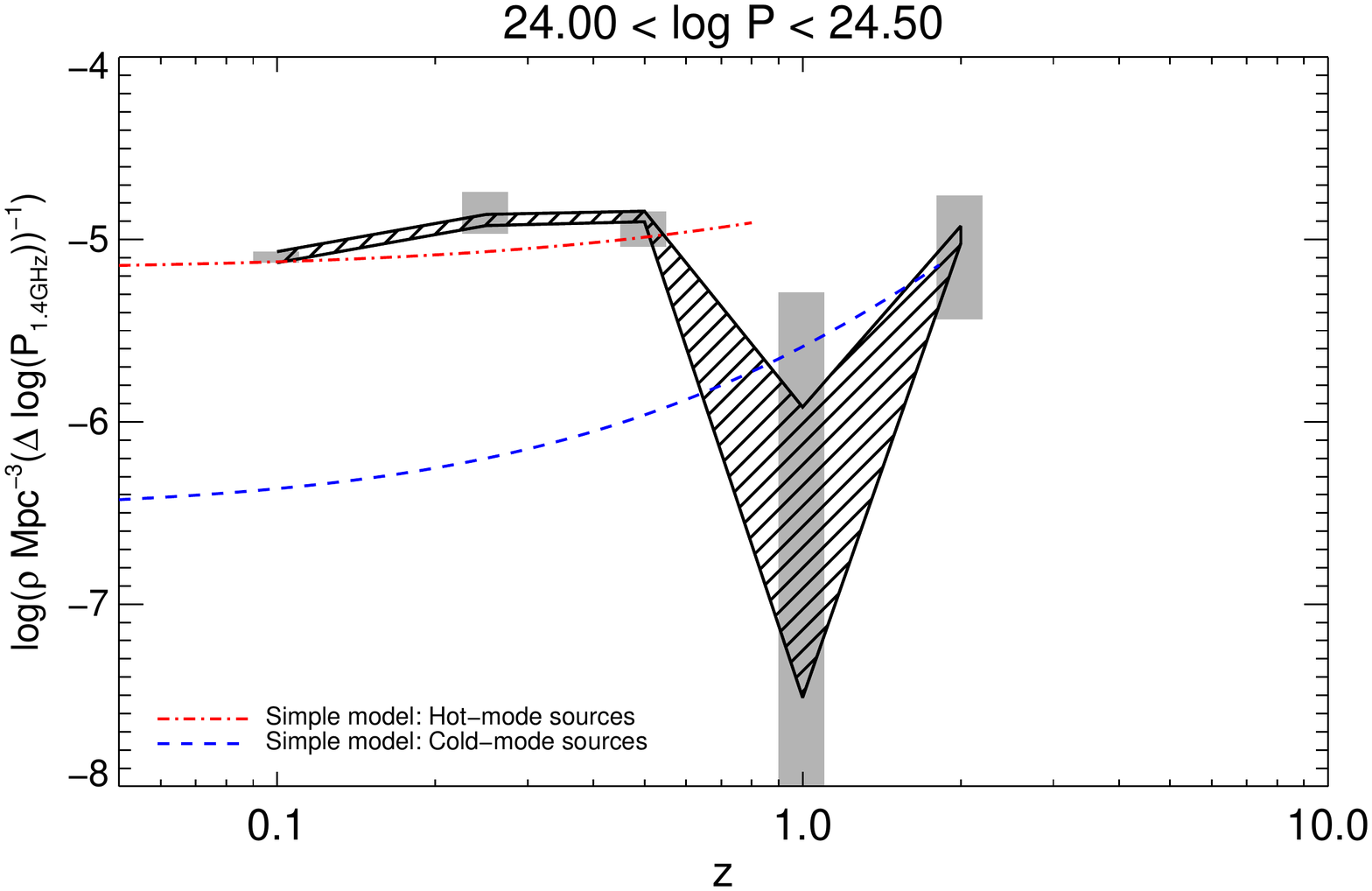}
\includegraphics[scale=0.3, trim=40 80 40 80, clip]{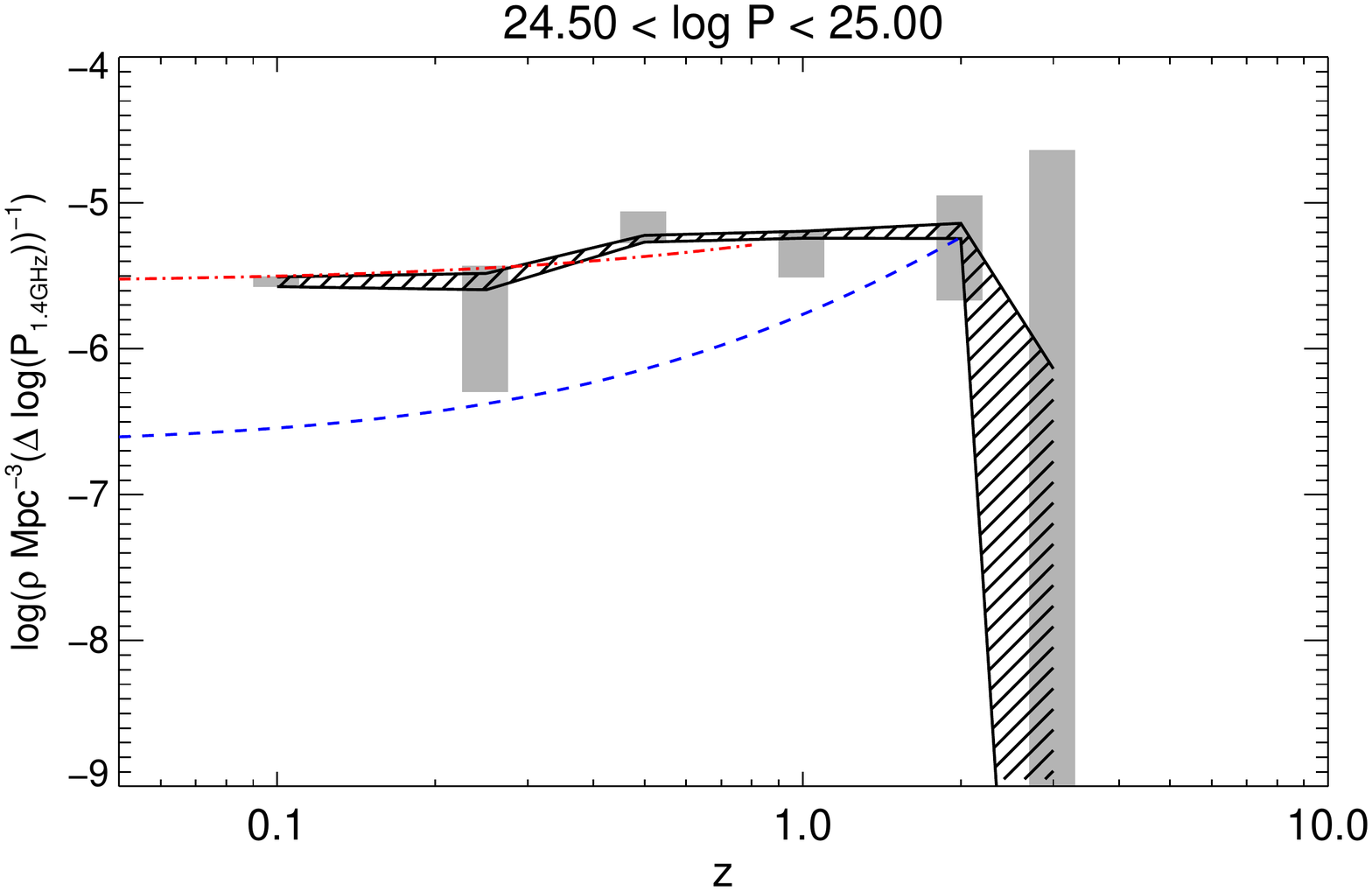}
\includegraphics[scale=0.3, trim=40 80 40 80, clip]{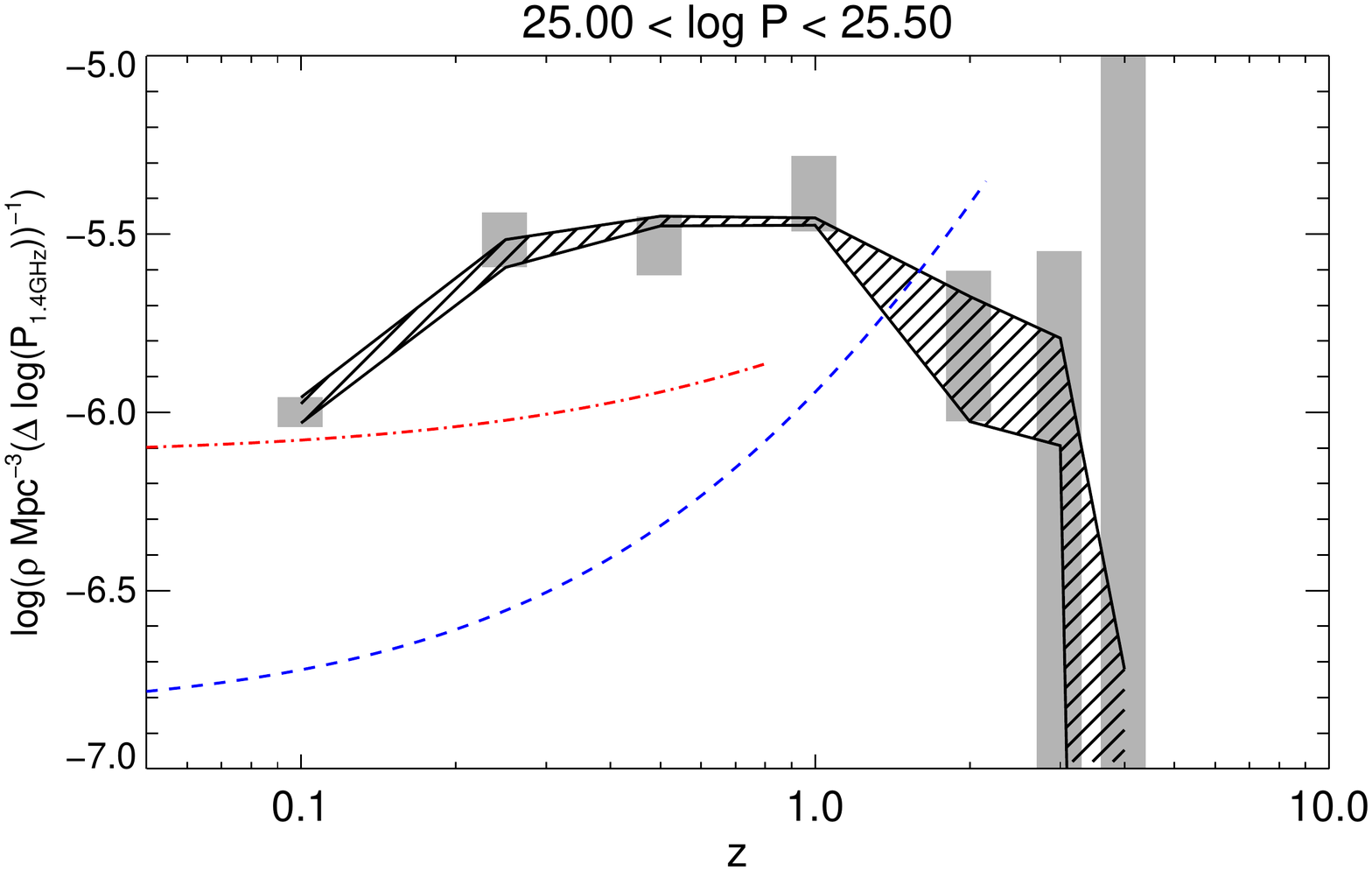}
\includegraphics[scale=0.3, trim=40 80 40 80, clip]{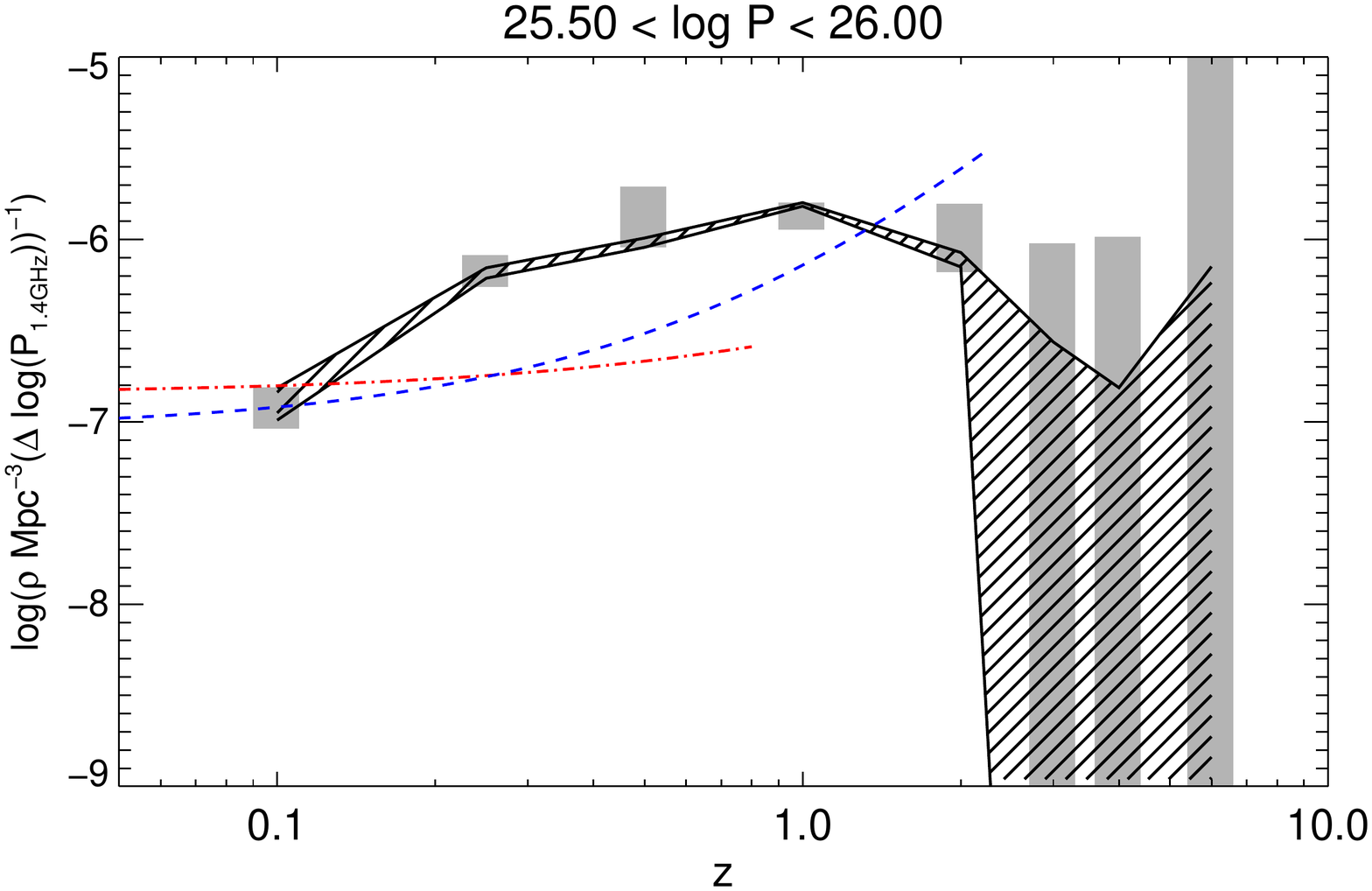}
\includegraphics[scale=0.3, trim=40 80 40 80, clip]{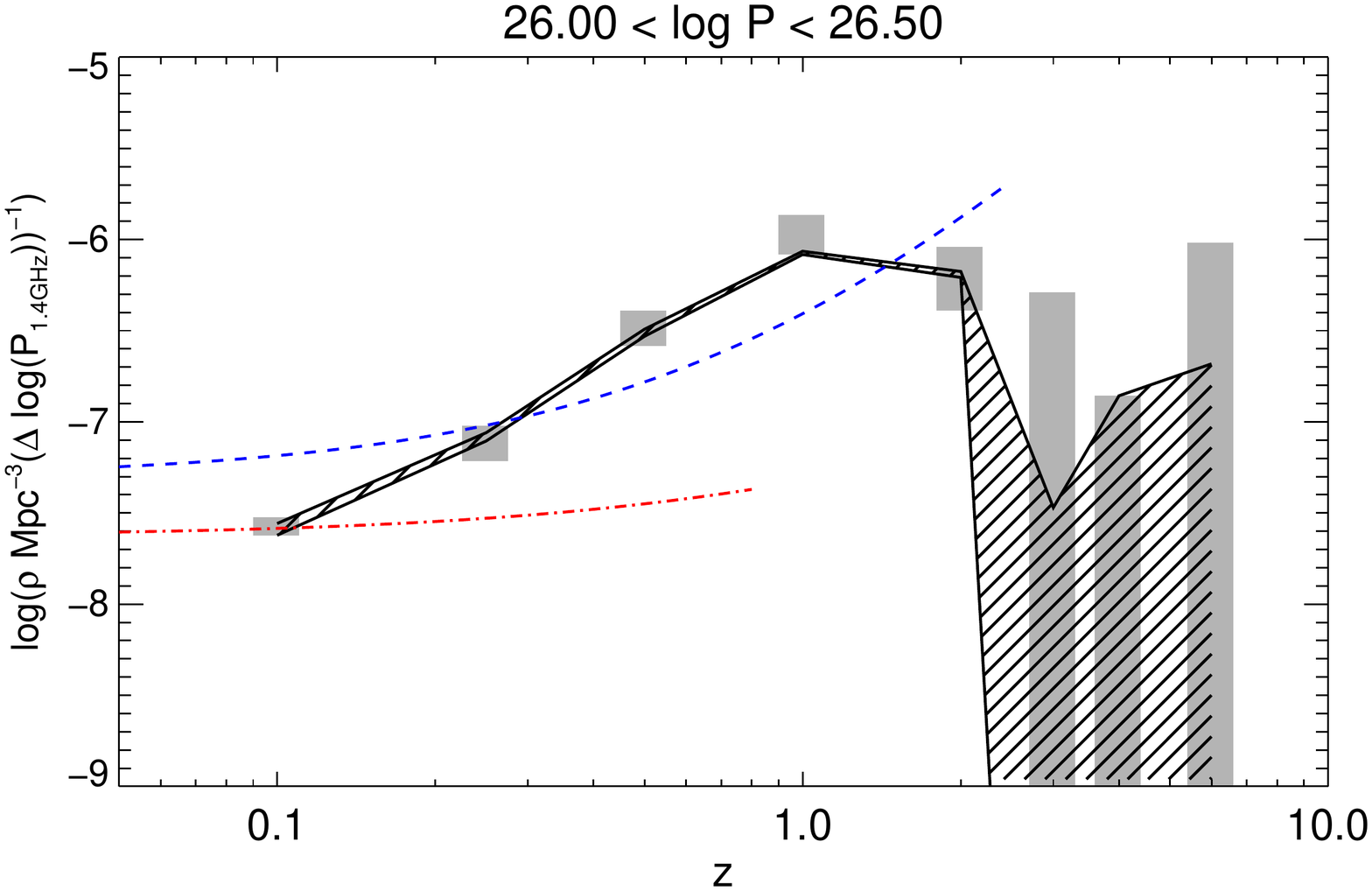}
\includegraphics[scale=0.3, trim=40 80 40 80, clip]{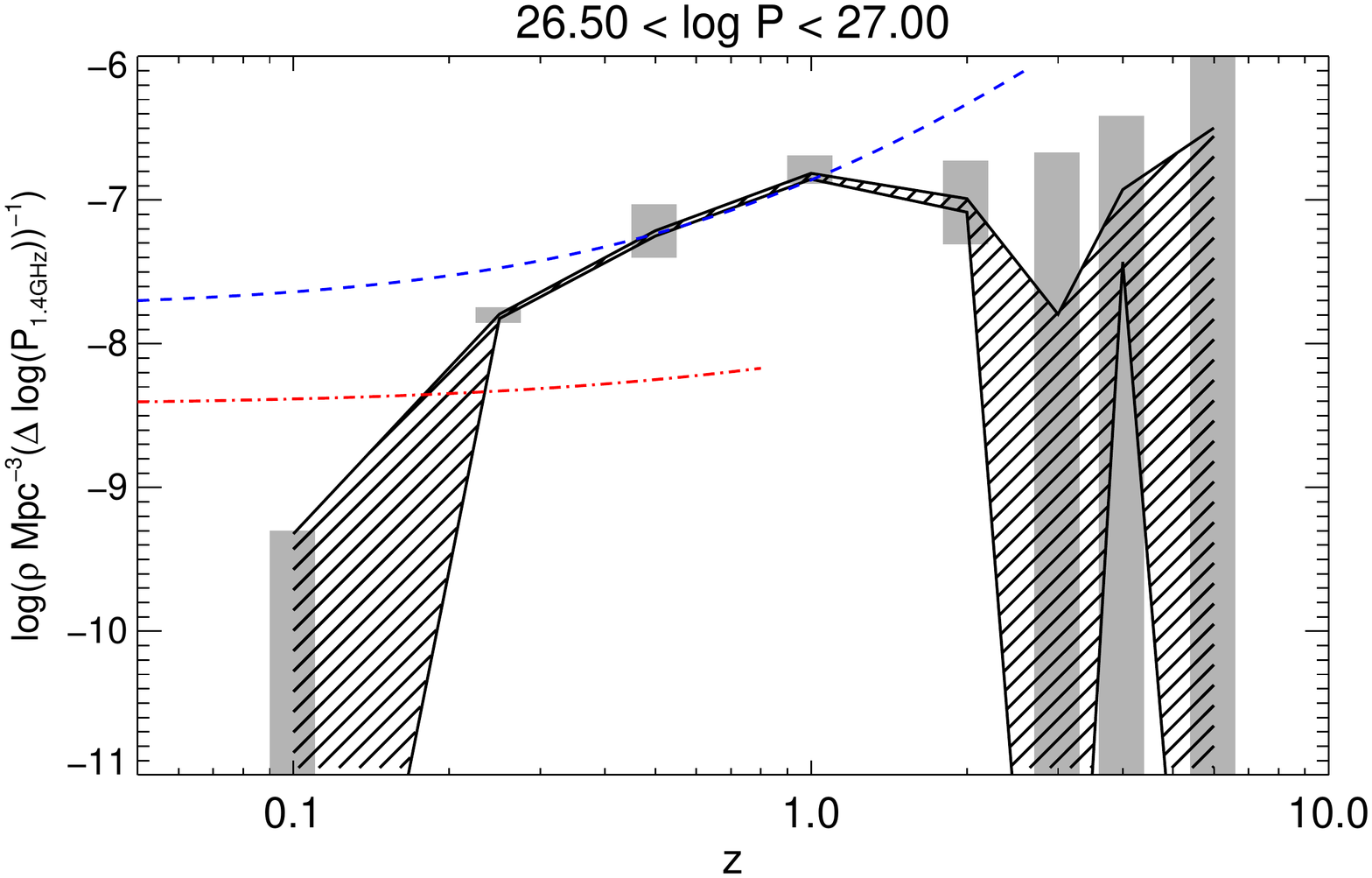}
\includegraphics[scale=0.3, trim=40 80 40 80, clip]{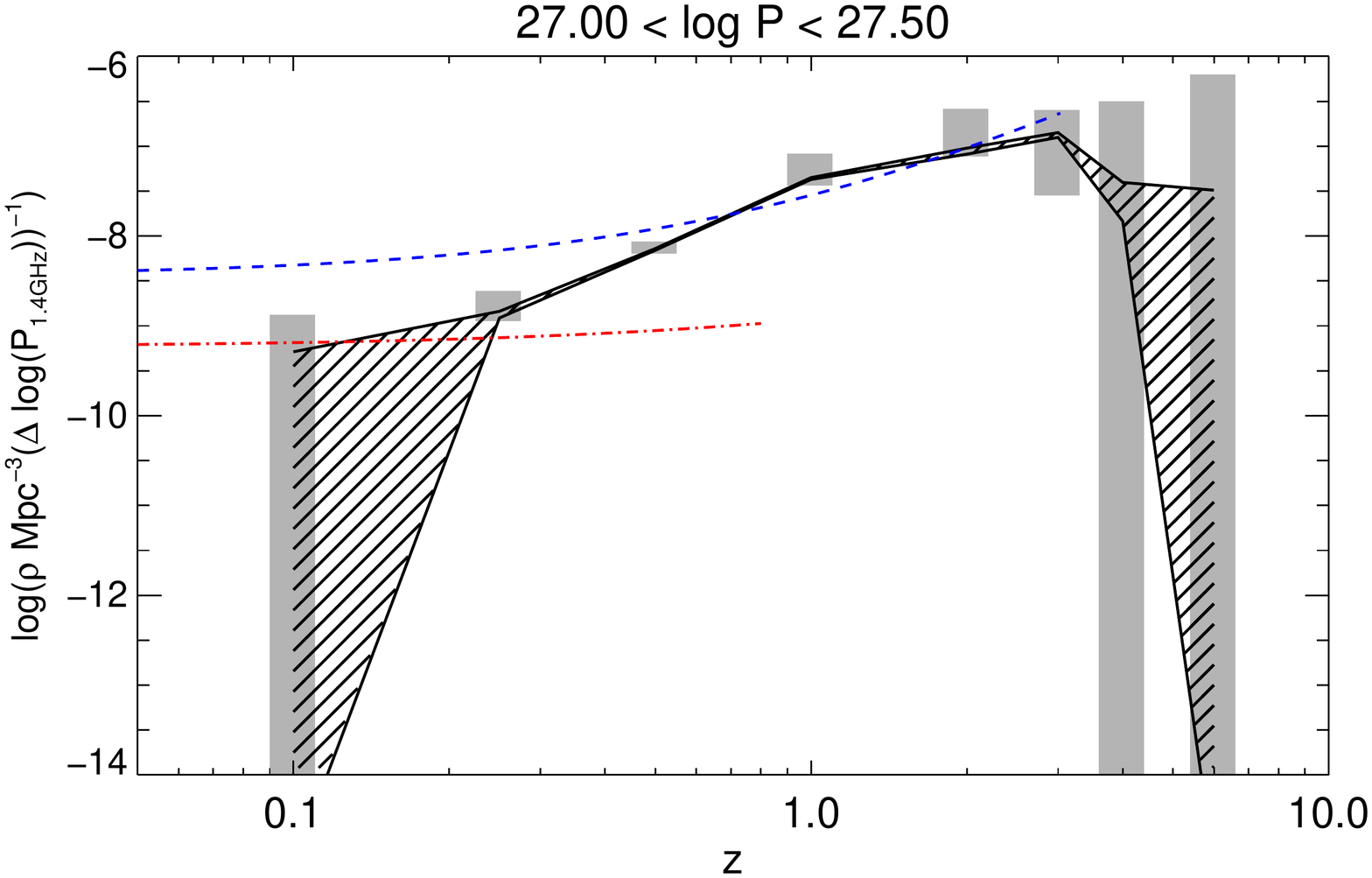}
\includegraphics[scale=0.3, trim=40 80 40 80, clip]{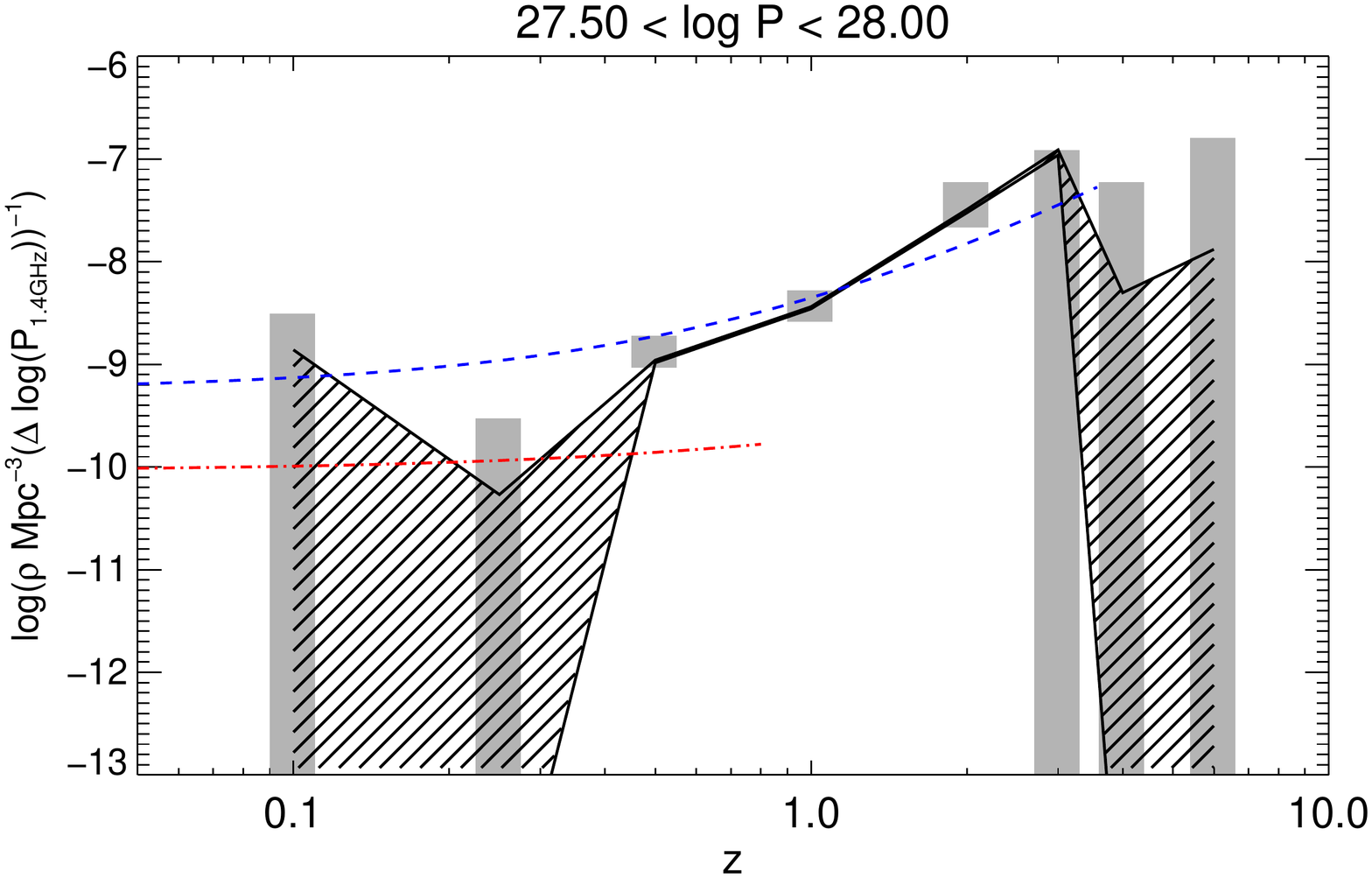}
\caption{\protect\label{rlfs} The model steep--spectrum RLFs, as a function of redshift, from the best--fitting $P$--$z$ grid (black shaded region). Results are only plotted if they are constrained by at least two of the input datasets (Figure \protect\ref{pz}). Dark grey shaded regions show the spread in results from varying the assumed value of the spectral index. Dotted and dashed lines show the predictions of the simple RLF model described in Section \protect\ref{toy_sec} for the hot--mode and cold--mode populations respectively (for clarity these are plotted up to the peak redshift value only). The brightest RLFs ($\log P > 26$) were fully constrained previously, and are therefore unchanged from R11. They are reproduced here for completeness and comparison with the simple RLF model.}
\end{figure*}

The predicted variation of space density with redshift for 8 radio power bins ($24 \leq \log P \leq 28$) is shown in Figure \ref{rlfs}. Note that the $\rho$ for a particular $P$--$z$ point is only shown if it is constrained by both the source counts and the redshift distributions (Figure \ref{pz}). 

The main benefit of including the SXDF sample in the modelling comes at the lowest powers ($\log P < 25.5$), where it allows the RLF behaviour to be studied at $z > 1$. Inspection of Figure \ref{rlfs} shows that decreases in space density are also present here, though these tend to occur at lower redshifts than for higher power ranges; this will be discussed further in Section \ref{z_turn}.

As part of the modelling process the original $P$--$z$ grid is converted into a grid of flux density and redshift to allow easier comparison with the source counts and redshift distributions. To make this $S$--$z$ grid a spectral index, $\alpha$, is needed; following R11 this is assumed to vary with redshift as $\alpha = 0.83 + 0.4 \log (1+z)$ \citep{ubachukwu}, with an additional uncertainty of $\alpha \pm 0.2$ at each redshift to account for any inter--bin variations. Figure \ref{rlfs} shows the range of best--fit results returned when this assumption is changed to either a constant, but very steep spectrum ($\alpha = 1.5$), a constant value of 0.8 (the mean value at low redshift), or a stronger increase with redshift ($\alpha = 0.8 + 0.25 z$). The general effect of using the steeper $\alpha$ values is to increase the overall density values, and weaken, though not remove, the $z \gtrsim 1$ space density turnovers as sources move to bins in the $S$--$z$ grid corresponding to higher radio powers. The assumption of the shallower value of $\alpha = 0.8$ has the opposite effect. The small spread in results seen at $26 < \log P < 26.5$ following this $\alpha$ variation suggests that this is the radio power bin which is most strongly constrained in the modelling process.

\section{The luminosity--dependence of the high--redshift turnover}
\label{z_turn}

\begin{figure}
\centering
\includegraphics[scale=0.37, trim=60 80 40 80, clip]{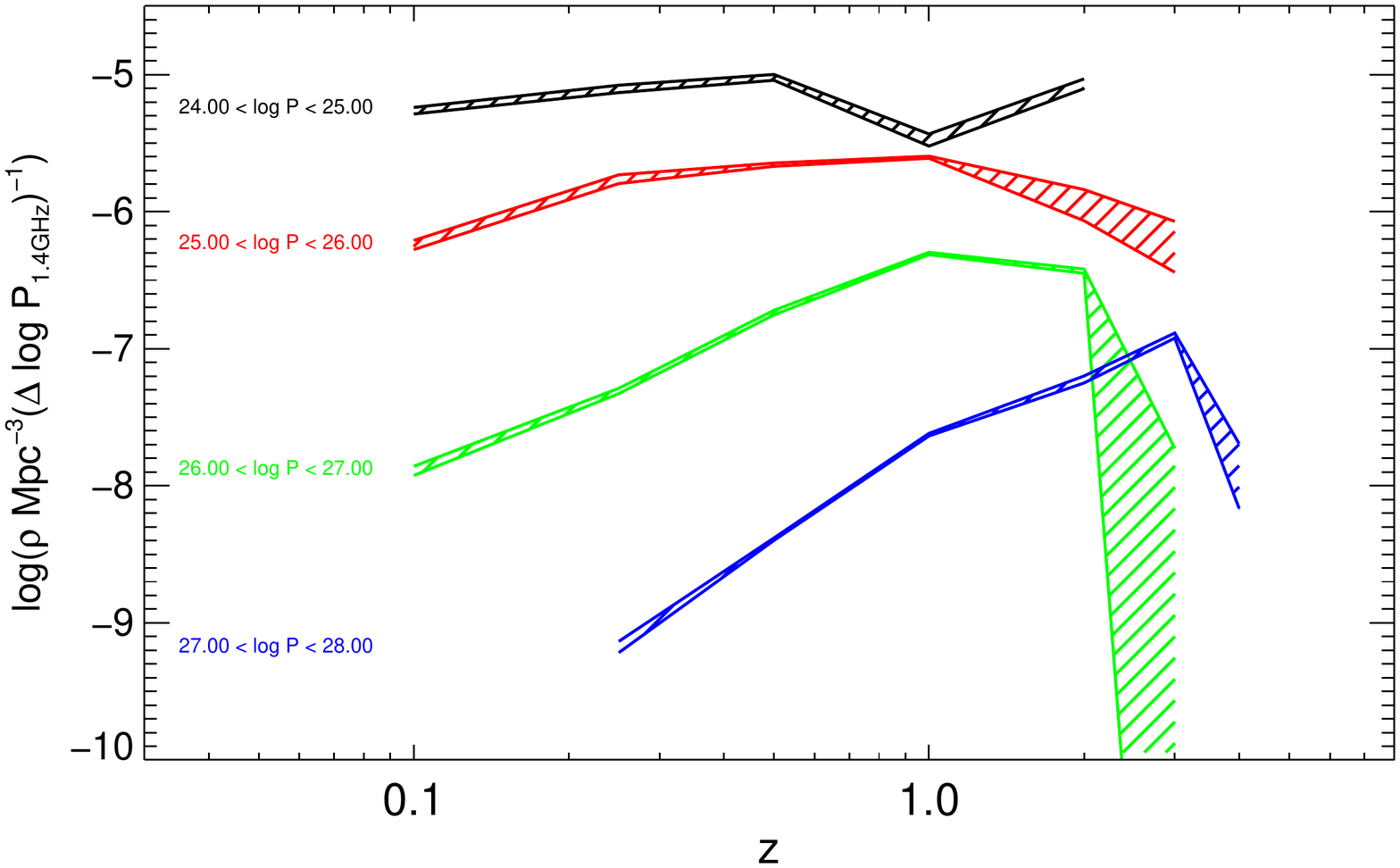}
\caption{\protect\label{av_rlf} The model steep--spectrum RLFs, as a function of redshift, from Figure \protect\ref{rlfs} plotted together to illustrate the changing position of the space density peak. Larger bins of $\Delta \log P = 1$ are used for clarity.}
\end{figure}

Inspection of Figure \ref{rlfs} suggests that, as in R11, the position of the space density peak is luminosity dependent. This can be clearly seen if the model steep--spectrum RLFs are shown together, as in Figure \ref{av_rlf}. The addition of the SXDF sample as a constraint to the RLF modelling allows this behaviour to be probed to lower radio powers than previously possible. 

\begin{figure}
\centering
\includegraphics[scale=0.37, trim=60 80 40 80, clip]{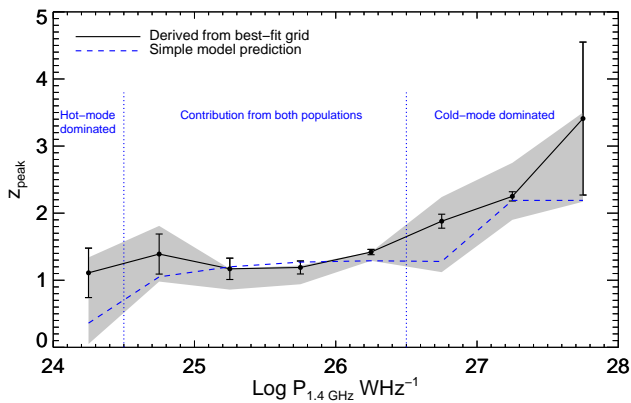}
\caption{\protect\label{zpeaks} \protect\label{toy} The variation in the redshift of the peak space density with radio power for the best--fitting steep spectrum grid determined here (black line), compared with that predicted using the simple model described in Section \ref{toy_sec} (blue line). Dotted lines delineate the relative contributions to the latter from the two underlying populations. The error bars show the uncertainty in the polynomial fits used to determine the peak position, whilst the shaded region represents the range in results found from varying the input parameters used in the RLF grid modelling. }
\end{figure}

The position of the peak redshift in each radio power range is determined by fitting a polynomial (generally of order 2, but of order 3 where necessary) to the model space density distributions in the $P$--$z$ grid. This overcomes the ambiguity arising from the discrete nature of the $P$--$z$ grid, and the wide redshift bins used. Figure \ref{zpeaks} shows the results of this for the best--fitting steep--spectrum grid presented in Section \ref{rlf_sec}, together with the spread arising from repeating this process for the grids determined using altered spectral indices. The small variation in values at $\log P = 26.25$ supports the previous identification of $26 < \log P < 26.5$ as the most constrained radio power bin in the model grid. 

The luminosity--dependence of the peak redshift, $z_{\rm peak}$, is clear, with the most powerful radio sources peaking at earlier times than their weaker counterparts. However, the extension to lower radio powers reveals an apparent flattening of the relationship: below $\log P \lesssim 26.5$ the results are consistent with a constant value of $z_{\rm peak}$.  

\subsection{Comparison with quasar behaviour}

The luminosity--dependent behaviour of the RLF peak redshift seen here for steep--spectrum radio sources has also been observed in other AGN populations \citep[e.g.][]{zotti09, hasinger05, richards05, delvecchio14}. A direct comparison of the $z_{\rm peak}$ changes is possible for flat--spectrum quasars, using the luminosity function (QLF) models of \citet{hopkins}; these are derived from a large set of observed QLFs, from the mid--infrared to X--ray. 

\begin{figure}
\centering
\includegraphics[scale=0.44, trim=60 80 40 80, clip]{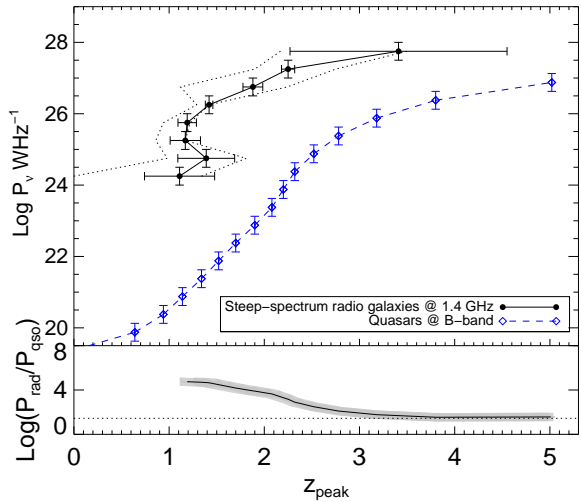}
\caption{\protect\label{zpeak_comp} A comparison of the variation of the redshift of the space density peak with radio luminosity (steep--spectrum sources; black solid line, with the uncertainty range arising from the assumed spectral index indicated by dotted lines) and B--band luminosity (quasars; blue dashed line). The subplot shows the ratio between the two luminosities as a function of $z_{\rm peak}$. The dotted line here indicates a constant ratio of $\log (P_{\rm 1.4 GHz} / P_{\rm B-band}) = 1.5$ - the minimum value historically adopted when classifying a source as radio--loud \citep{kellermann89}.
}
\end{figure}

The optical B--band (0.44 $\mu$m) is chosen for extracting the quasar behaviour, for ease of comparison to the luminosities in the radio waveband. It should be noted however that the variation in peak redshift calculated using the bolometric QLF instead is qualitatively similar. The Hopkins et al. model is queried for a luminosity range over which it is reliable ($20 < \log P_{\rm B-band} < 27$), and the $z_{\rm peak}$ value determined in each case. The subsequent result is shown in Figure \ref{zpeak_comp}, together with that already found for the steep--spectrum sources. The ratio of the two curves is also shown, as a function of peak redshift. 

The shape of the $z_{\rm peak}$--$\log P_{B-band}$ curve derived for the quasar population is broadly similar to that of the radio galaxies, with indications of flattening at $24 < \log P_{\nu} < 26$, before continuing to decrease at lower powers. The $\log (P_{\rm 1.4 GHz} / P_{\rm B-band})$ ratio begins to increase from a constant value of $\sim 1.5$ at $z > 3$, within the uncertainties; it is interesting to note that $\log (P_{\rm 1.4 GHz} / P_{\rm B-band}) \geq 1.5$ is the historical definition of a radio--loud quasar \citep{kellermann89}.

\subsection{The relation to cosmic downsizing}
\label{toy_sec}
The obvious physical interpretation of the luminosity--dependent space density peak seen here is cosmic downsizing, in which the most massive black holes have formed by $z \sim 4$, with the less massive forming at later times. This apparently contradictory anti--hierarchical growth can be explained if the dominant mechanism of AGN fuelling changes with cosmic time from cold gas accretion via major mergers (`cold--mode') to radiatively inefficient accretion directly from hot gas haloes (`jet--mode' or `hot--mode'). Simulations of AGN evolution support this picture; they show that hierarchical black hole growth is able to produce the observed downsizing trend if inefficient accretors come to dominate at low redshift \citep{fanidakis12, hirschmann13}. 

\begin{table*}
\caption{\protect\label{table} The peak redshift ($z_{\rm peak}$) values used in the simple model of hot and cold--mode RLF evolution, described in Section \protect\ref{toy_sec}. In this model hot--mode sources increase as $(1+z)$ to $z_{\rm peak} = 0.8$, whereas cold--mode sources have a luminosity dependent value of $z_{\rm peak}$ and a pre--peak increase of $(1+z)^{3}$. Both populations have a post--peak decline of $(1+z)^{-6.5}$. The numbers in brackets are the peak values after the predicted RLFs are re--sampled using the same redshift bins as in the grid modelling. The `Hot+Cold Mode' column gives $z_{\rm peak}$ for the combined hot and cold--mode RLFs, extracted via polynomial fitting. The `Grid Model' column gives the corresponding $z_{\rm peak}$ determined from the best--fit model grid (Section \protect\ref{rlf_sec}); the lower and upper bounds give the range arising from varying the input parameters used in the grid modelling, as shown by the shaded region in Figure \protect\ref{zpeaks}.}
\centering
\begin{tabular}{ccccc}
\hline\hline
Radio power range    & \multicolumn{4}{c}{$z_{\rm peak}$}     \\
                     &  Hot--mode & Cold--mode  & Hot+Cold mode & Grid model\\
\hline                                                                           
24.00 -- 24.50       &   0.8 (0.5)&  1.8 (1.0)  & 0.4           & 1.1$^{0.2}_{1.1}$       \\
24.50 -- 25.00       &   0.8 (0.5)&  2.0 (2.0)  & 1.1           & 1.4$^{0.4}_{0.4}$       \\
25.00 -- 25.50       &   0.8 (0.5)&  2.2 (2.0)  & 1.2           & 1.2$^{0.0}_{0.3}$       \\
25.50 -- 26.00       &   0.8 (0.5)&  2.3 (2.0)  & 1.3           & 1.2$^{0.0}_{0.3}$       \\
26.00 -- 26.50       &   0.8 (0.5)&  2.4 (2.0)  & 1.3           & 1.4$^{0.0}_{0.1}$       \\
26.50 -- 27.00       &   0.8 (0.5)&  2.7 (2.0)  & 1.3           & 1.9$^{0.4}_{0.8}$       \\
27.00 -- 27.50       &   0.8 (0.5)&  3.0 (3.0)  & 2.2           & 2.3$^{0.5}_{0.4}$       \\
27.50 -- 28.00       &   0.8 (0.5)&  3.6 (3.0)  & 2.2           & 3.4$^{0.1}_{1.2}$       \\
\hline
\end{tabular}
\end{table*}

Locally, hot--mode sources make up the majority of the radio--galaxy population at $\log P_{\rm 1.4 GHz} \lesssim 26$ \citep{best2012}, though both types are generally found over the range of radio powers studied. In comparison, by $z \sim 0.7$ the space densities of the two are similar at low radio powers, with the cold--mode sources becoming more dominant at $\log P_{\rm 1.4 GHz} \gtrsim 26$ \citep{best14}. The flattening seen here in the $z_{\rm peak}$ vs. $\log P_{\rm 1.4 GHz}$ relation occurs at corresponding radio power to this transition, lending further support to the assertion that the RLF evolution seen here directly represents the interplay between the two populations. It also suggests the weaker, likely hot--mode, sources undergo little evolution in space density; a finding broadly replicated in \citet{best14} which directly catalogued sources into the two classes.

This assertion can be tested by predicting the dual population space density evolution, taking as a starting point the well--defined local RLFs for the two classes \citep{best2012}. For the evolution of the hot--mode population, we follow \citet{best14} in assuming that their evolution broadly follows that of the space density of massive quiescent galaxies, which constitute their host galaxies. \citet{best14} showed that the space density of these galaxies remains relatively flat out to $z \sim 0.8$, and then rapidly decreases as $(1+z)^{-6.5}$. We modify this to allow for an increase in the space density as $(1+z)$ out to the peak redshift at $z=0.8$, as this is a better match to the evolution of the $10^{24} < P_{\rm 1.4 GHz} < 10^{25}$ RLFs in the model grid. 

For the cold--mode population, we determine the (luminosity--dependent) peak redshift by comparison with the optical quasars. Figure \ref{zpeak_comp} suggests that there is a roughly constant ratio between the luminosities of radio galaxies and quasars at high values of $z_{\rm peak}$, where the cold--mode sources dominate. Assuming this is true for this population, $z_{\rm peak}$ for a particular radio luminosity is therefore taken as the peak of the corresponding $B$--band quasar luminosity. For this population, we allow a more rapid pre--peak increase in space density of $(1+z)^{3}$, which matches the evolution seen at the highest radio luminosities in the model grid. The post--peak decrease is again taken as $(1+z)^{-6.5}$.

The predicted RLFs for the two populations are then combined, sampled using the same luminosity and redshift bins as in the model grid, and the peak redshift extracted. 

The variation in peak redshift with radio luminosity arising from this simple model is shown in Figure \ref{toy}. It does well at reproducing the overall behaviour derived from the best--fit model grid. At the faintest ($\log P_{\rm 1.4 GHz} < 24.5$) luminosities the predicted RLFs are dominated by the hot--mode sources, whilst the cold--mode sources dominate at $\log P_{\rm 1.4 GHz} > 26.5$. In-between, the RLFs are a combination of the two classes. This is reflected in the changing shape of the $z_{\rm peak}$ -- $\log P_{\rm 1.4 GHz}$ curve: the plateau at $24.5 < \log P_{\rm 1.4 GHz} < 26.5$ arises, in this scenario, from the transition from the cold--mode to the hot--mode populations.

\section{Summary \& Conclusions}
\label{summ}

The results presented in this paper have demonstrated the benefits of extending the $P$--$z$ plane coverage to lower radio powers when measuring the evolution of the steep--spectrum RLF. The addition of the updated SXDF sample to the constraining datasets used in the modelling process shows that the high--redshift turnover in the RLF persists to $\log P_{\rm 1.4 GHz} > 24$ - an order of magnitude fainter than could be measured previously.

The redshift at which the peak space density occurs is found to be luminosity--dependent at $\log P_{\rm 1.4 GHz} \gtrsim 26$, with the most powerful sources peaking at earlier times than their weaker counterparts. This mirrors that seen when considering quasar optical luminosity functions, which is to be expected as the radio population is dominated by cold--mode (i.e. quasar--like) sources at these luminosities. The presence of a mean ratio for these powers ($\log[P_{\rm 1.4 GHz} / P_{\rm B-band}] \sim 1.5$) which remains constant with $z_{\rm peak}$ again suggests that the radio and optical observations are seeing the same objects. This picture changes at lower radio luminosities ($\log P_{\rm 1.4 GHz} \lesssim 26$). Here the value of the peak redshift remains constant within the uncertainties, whereas it continues to fall for the optical population. This implies an increase in the $P_{\rm 1.4 GHz} / P_{\rm B-band}$ ratio and suggests that the dominant radio population is changing from cold--mode to hot--mode. 

Simple models for the space density evolution of the hot--mode and cold--mode populations are able to reproduce the observed luminosity dependence in the peak redshift, reinforcing the conclusion that this behaviour arises from the transition between the two populations. To properly understand the relative contributions of the hot--mode and cold--mode sources to the RLF evolution, their behaviour should be minimised separately in the fitting. This is trivial to do with this grid--based modelling method, but the input data needed are currently lacking across the required redshift range. Future large, deep, radio surveys in regions with extensive multiwavelength coverage will make this possible, whilst also allowing the coverage of the $P$--$z$ plane to extend to radio powers of $\log P_{\rm 1.4 GHz} < 24$. 

\section*{Acknowledgments}

EER acknowledges financial support from NWO (grant number: NWO-TOP LOFAR 614.001.006). EER and JA thank the Leiden/ESA astrophysics program for summer students (LEAPS) which supported JA in Leiden. The authors would like to thank Chris Simpson for sharing the $q$ parameters in the SXDF catalogue.

\setlength{\bibhang}{\parindent}
\setlength\labelwidth{0.0em}

\label{lastpage}


\begin{thebibliography}{}

\bibitem[Best et al.(2003)]{CENSORS1} Best, P.~N., Arts, J.~N., R{\"o}ttgering, H.~J.~A., et al.\ 2003, \mnras, 346, 627 

\bibitem[\protect\citeauthoryear{{Best}, {Kaiser}, {Heckman} \&
  {Kauffmann}}{{Best} et~al.}{2006}]{best06}
{Best} P.~N.,  {Kaiser} C.~R.,  {Heckman} T.~M.,    {Kauffmann} G.,  2006,
  MNRAS, 368, L67

\bibitem[Best et al.(2007)]{best07} Best, P.~N., von der Linden, A., Kauffmann, G., Heckman, T.~M., \& Kaiser, C.~R.\ 2007, \mnras, 379, 894 

\bibitem[Best \& Heckman(2012)]{best2012} Best, P.~N., \& Heckman, T.~M.\ 2012, \mnras, 421, 1569 

\bibitem[Best et al.(2014)]{best14} Best, P.~N., Ker, L.~M., Simpson, C., Rigby, E.~E., \& Sabater, J.\ 2014, \mnras, 445, 955 

\bibitem[\protect\citeauthoryear{Bower et al.}{2006}]{bower06} Bower R. G., Benson A. J., Malbon R., Helly J. C., Frenk C. S., Baugh C. M., Cole S., Lacey C. G., 2006, MNRAS, 370, 645

\bibitem[Croton et al.(2006)]{croton06} Croton, D.~J., et al.\ 2006, \mnras, 365, 11 

\bibitem[Delvecchio et al.(2014)]{delvecchio14} Delvecchio, I., Gruppioni, C., Pozzi, F., et al.\ 2014, \mnras, 360 

\bibitem[De Zotti et al.(2010)]{zotti09} De Zotti, G., Massardi, M., Negrello, M., \& Wall, J.\ 2010, \aapr, 18, 1 

\bibitem[Donley et al.(2005)]{donley05} Donley, J.~L., Rieke, G.~H., Rigby, J.~R., \& P{\'e}rez-Gonz{\'a}lez, P.~G.\ 2005, \apj, 634, 169 

\bibitem[\protect\citeauthoryear{{Dunlop} \& {Peacock}}{1990}]{DP90}{Dunlop} J.~S.,  {Peacock} J.~A.,  1990, \mnras, 247, 19

\bibitem[\protect\citeauthoryear{Fabian et al.}{2006}]{fabian06} Fabian, A.~C., Sanders, J.~S., Taylor, G.~B., et al.\ 2006, \mnras, 366, 417

\bibitem[Fanidakis et al.(2012)]{fanidakis12} Fanidakis, N., Baugh, C.~M., Benson, A.~J., et al.\ 2012, \mnras, 419, 2797 

\bibitem[Hasinger et al.(2005)]{hasinger05} Hasinger, G., Miyaji, T., \& Schmidt, M.\ 2005, \aap, 441, 417 

\bibitem[Hirschmann et al.(2012)]{hirschmann12} Hirschmann, M., Somerville, R.~S., Naab, T., \& Burkert, A.\ 2012, \mnras, 426, 237 

\bibitem[Hirschmann et al.(2014)]{hirschmann13} Hirschmann, M., Dolag, K., Saro, A., et al.\ 2014, \mnras, 442, 2304 

\bibitem[Hopkins et al.(2007)]{hopkins} Hopkins, P.~F., Richards, G.~T., \& Hernquist, L.\ 2007, \apj, 654, 731 

\bibitem[\protect\citeauthoryear{{Jarvis}, {Rawlings}, {Willott}, {Blundell},
  {Eales} \& {Lacy}}{{Jarvis} et~al.}{2001}]{jarvis01b}
{Jarvis} M.~J.,  {Rawlings} S.,  {Willott} C.~J.,  {Blundell} K.~M.,  {Eales}
  S.,    {Lacy} M.,  2001, \mnras, 327, 907

\bibitem[Kellermann et al.(1989)]{kellermann89} Kellermann, K.~I., Sramek, R., Schmidt, M., Shaffer, D.~B., \& Green, R.\ 1989, \aj, 98, 1195 

\bibitem[Ibar et al.(2008)]{ibar08} Ibar, E., Cirasuolo, M., Ivison, R., et al.\ 2008, \mnras, 386, 953 

\bibitem[Lilly et al.(2009)]{lilly09} Lilly, S.~J., Le Brun, V., Maier, C., et al.\ 2009, \apjs, 184, 218 

\bibitem[McAlpine et al.(2013)]{mcalpine13} McAlpine, K., Jarvis, M.~J., \& Bonfield, D.~G.\ 2013, \mnras, 436, 1084 

\bibitem[\protect\citeauthoryear{{Nelder} \& {Mead}}{{Nelder} \& {Mead}}{1965}]{downhillsimplex}{Nelder} J.~A.,  {Mead} R.,  1965, Computer Journal, 7, 308

\bibitem[Norris et al.(2013)]{norris} Norris, R.~P., Afonso, J., Bacon, D., et al.\ 2013, \pasa, 30, 20 

\bibitem[Richards et al.(2005)]{richards05} Richards, G.~T., et al.\ 2005, \mnras, 360, 839 

\bibitem[Rigby et al.(2011)]{paper1} Rigby, E.~E., Best, P.~N., Brookes, M.~H., et al.\ 2011, \mnras, 416, 1900 

\bibitem[Sadler et al.(2002)]{Sadler02} Sadler, E.~M., et al.\ 2002, \mnras, 329, 227 

\bibitem[Schinnerer et al.(2010)]{schinnerer} Schinnerer, E., Sargent, M.~T., Bondi, M., et al.\ 2010, \apjs, 188, 384 

\bibitem[\protect\citeauthoryear{{Shaver}, {Wall}, {Kellermann}, {Jackson} \&
  {Hawkins}}{{Shaver} et~al.}{1996}]{shaver}{Shaver} P.~A.,  {Wall} J.~V.,  {Kellermann} K.~I.,  {Jackson} C.~A., {Hawkins} M.~R.~S.,  1996, Nature, 384, 439

\bibitem[Simpson et al.(2006)]{simpson06} Simpson, C., Mart{\'{\i}}nez-Sansigre, A., Rawlings, S., et al.\ 2006, \mnras, 372, 741 

\bibitem[Simpson et al.(2012)]{simpson12} Simpson, C., Rawlings, S., Ivison, R., et al.\ 2012, \mnras, 421, 3060 

\bibitem[Smolcic et al.(2015)]{smolcic15} Smol{\v c}i{\'c}, V., Padovani, P., Delhaize, J., et al.\ 2015, arXiv:1501.04820 

\bibitem[Smol{\v c}i{\'c} et al.(2008)]{smolcic} Smol{\v c}i{\'c}, V., et al.\ 2008, \apjs, 177, 14 

\bibitem[\protect\citeauthoryear{Ubachukwu}{Ubachukwu et al.}{1996}]{ubachukwu} Ubachukwu, A. A., Ugwoke, A. C., \& Ogwo, J. N., 1996, \apss, 238, 151U

\bibitem[\protect\citeauthoryear{{Waddington}, {Dunlop}, {Peacock} \& {Windhorst}}{{Waddington} et~al.}{2001}]{waddington01}{Waddington} I.,  {Dunlop} J.~S.,  {Peacock} J.~A.,    {Windhorst} R.~A.,
  2001, \mnras, 328, 882

\end{thebibliography}
\end{document}